\documentclass[twocolumn,twocolappendix]{aastex631}
\usepackage[utf8]{inputenc}
\usepackage{amsmath, mathtools}
\usepackage{graphicx}
\usepackage{wrapfig}
\usepackage{float}
\usepackage{natbib,color,ulem,hyperref}

\makeatletter
\renewcommand\paragraph{\@startsection{paragraph}{4}{\z@}%
            {-2.5ex\@plus -1ex \@minus -.25ex}%
            {1.25ex \@plus .25ex}%
            {\normalfont\normalsize}}
\makeatother
\newcommand{\y}{\color{red}}
\newcommand{\z}{\color{orange}}
\newcommand{\w}[1]{#1} 

\newcommand{\be}{\begin{eqnarray}}
\newcommand{\ee}{\end{eqnarray}}
\newcommand{\beqn}{\begin{eqnarray}}
\newcommand{\eeqn}{\end{eqnarray}}
\newcommand{\bi}{\begin{itemize}}
\newcommand{\ei}{\end{itemize}}

\def\g{\, \rm g}
\def\K{\, \rm K}

\def\s{\, \rm s}
\def\km{\, \rm km}

\def\m{\, \rm m}

\def\apjs{\, ApJS}
\def\aap{\, A\&Ap}

\def\apjl{\, ApJL}

\usepackage{ulem}
\newcommand{\x}{\sout}

\shorttitle{Binary Eccentricities}
\shortauthors{Wu et al.}

\begin{document}

\title[Why are Stellar Binaries Eccentric?]{Eccentricities of Close Stellar Binaries}

\author[0000-0003-0511-0893]{Yanqin Wu}
\affiliation{Department of Astronomy \& Astrophysics, University of  Toronto, 50 St George Street, Toronto, ON M5S\,3H4, Canada}
\author[0000-0002-1032-0783]{Sam Hadden}
\affiliation{Canadian Institute of Theoretical Astrophysics, University of Toronto, 50 St George Street, Toronto, On M5S\,3H4, Canada}
\author[0000-0001-9420-5194]{Janosz Dewberry} 
\affiliation{Canadian Institute of Theoretical Astrophysics, University of Toronto, 50 St George Street, Toronto, On M5S\,3H4, Canada}
\author[0000-0002-6871-1752]{Kareem El-Badry}
\affiliation{Department of Astronomy, California Institute of Technology, 1200 E. California Blvd., Pasadena, CA 91125, USA}
\author[0000-0001-9732-2281]{Christopher D. Matzner}
\affiliation{Department of Astronomy \& Astrophysics, University of  Toronto, 50 St George Street, Toronto, ON M5S\,3H4, Canada}










\begin{abstract}
Orbits of stellar binaries are in general eccentric. {\w These eccentricities encode information about their early lives.}  
Here, we use thousands of
main-sequence binaries from the Gaia DR3 catalog to reveal that, binaries inwards of a few AU exhibit a simple Rayleigh distribution with a mode $\sigma_e  \simeq 0.3$. We find the same distribution for binaries from M to A spectral types, and from tens of days to $10^3$days (possibly extending to tens of AU).
This observed distribution is most likely primordial {\w and its invariance suggests a single universal process.  One possibility is eccentricity excitation by circumbinary disks. Another, as is suggested by the Rayleigh form, is weak scattering and ejection of brown-dwarf objects. We explore this latter scenario and find that the binary eccentricities reach an equi-partition value of $\sigma_e \simeq \sqrt{M_{\rm bd}/M_*}$. 
So to explain the observed mode, the brown dwarfs will have to be of order one tenth the stellar masses, and be at least as abundant in the Galaxy as the close binaries. The veracity of {\w both} proposals remains to be tested. }
\end{abstract}





\section{Introduction}\label{sec1}

Binary stars are common in the Galaxy \citep{DM91,Fischer1992}.
Their orbits are invariably eccentric. The distribution of these eccentricities encodes accessible information about their formation. {\w It is also used in} a wide range of studies. 
Surprisingly, for binaries closer than tens of AU (`close binaries'),  such a fundamental property 
{\w has not been definitely determined}. 


Multiple lines of evidence suggest two main modes of binary formation {\w exist} \citep[see][for reviews]{Duchene2013,offner2023}. Wide binaries \citep[$> 50$ AU,\footnote{Interestingly, $50$AU, or $P\sim 10^5$days, is roughly the peak of the binary distribution for Sun-like stars \citep{DM91,Raghavan2010}.} see, e.g.,][]{Parker2009}
likely form following direct collapse of individual components. These are either weakly bound at birth or are captured after birth.  Their occurrence  is insensitive to or rises with stellar metallicity  \citep{elbadry_rix_19,Hwang2021},
and the two components appear to be randomly paired in mass \citep{Moe2017}.
Close binaries, on the other hand, are now thought to have formed through gravitational fragmentation in massive circum-stellar disks \citep[see review by][]{kratterlodato16}. Their occurrence anti-correlates with stellar metallicity \citep{Moe2019,Mazzola2020}, and the two masses are correlated \citep{Raghavan2010,Duchene2013,Moe2017} with an excess of equal-mass `twin-binaries' \citep[][]{Moe2017,Elbadrytwin}.
 
%

The different formation pathways also manifest in binary eccentricities -- a property that is  observationally accessible and  dynamically informative. Wide binaries are observationally determined to have a thermal distribution ($dN/de \propto e$), as inferred {\w using} spectroscopic 
\citep{DM91,Raghavan2010} and visual binaries \citep{Tokovinin2020,Hwang2022}.\footnote{Very wide binaries ($>10^3$ AU) also appear to be super-thermal \citep{Tokovinin2020,Hwang2022}, suggesting another mechanism at play.} This reflects a  dynamic past in the birth clusters \citep{Parker2009}, where 
plentiful scatterings with other stars have relaxed the binaries towards a `thermal' equilibrium \citep{Jeans1919,Heggie1975}.

Such a concordance between theory and observation does not, however, extend to close binaries. There are no theoretical predictions for their e-distribution, owing to the uncertain fragmentation process. On the observational side, their e-distribution remains murky, though it clearly differs from that of the wide binaries. 
The current wisdom is that they are consistent with being ``uniform'',\footnote{Binaries with massive primaries (O/B stars) appear `thermal', down to periods as short as tens of days, possibly related to their higher triple fraction
\citep{Moe2017}. 
} $dN/de \propto e^0$ \citep{Tokovinin2000,Raghavan2010,Duchene2013,Moe2017,Geller2021,Hwang2022}.
{\w This conclusion, besides from having no ready theoretical explanation, is drawn from limited samples.} For instance, if we restrict to the period range $10^2-10^3$d (see below), 
we are left with only $18$ binaries in the volume-complete (to 25\,pc), Sun-like sample from \citet{Raghavan2010}.

{\w Larger samples do exist. The survey of nearby G-dwarfs by \citet{Udry1998} returns $N=110$ such systems; the massive compendium of all spectroscopic binaries, SB9 \citep{SB9}, contains nearly a thousand similar systems;
and the APOGEE survey constrained eccentricities for $\sim 8600$ FGK binaries beyond $100$ days \citep{PriceWhelan2020}. However, although results from these surveys hint at an eccentricity distribution that is  not ``uniform'', there have been no firm conclusions, largely because the selection bias on eccentricity has not been properly modeled. }
%
{\w It is clear that a binary sample, with a well calibrated completeness in eccentricity,} is sorely needed.

The Gaia mission \citep{gaia2016,gaia2023}, especially with its most recent non-single-star catalogue published in DR3 \citep{Arenou2023}, provides just this sample. The Gaia binaries, as we show here, sharpen our vision dramatically.  The e-distribution for those inward of a few AU follows, distinctly, a Rayleigh distribution. 



\section{GAIA Binaries }\label{sec:observation}

Here we will study Gaia binaries with  eccentricities explicitly determined by astrometry and/or radial velocities. 
The details of our binary selection are in Appendix \ref{sec:A1}, and a short form is presented in Table \ref{tab:sample}. The `full' main-sequence binary sample includes some $150,000$ systems. We pare down this large set by different cuts (Table \ref{tab:sample}) and study their respective properties. Among these, we highlight results from the so-called `gold' sample,  $\sim 3000$ Sun-like binaries that have periods from $10^2-10^3$ days, 
and that lie within $150$\,pc.  Binaries with too short a period can be affected by tidal circularization, while DR3 extends to just beyond $10^3$ days; binaries beyond $150$\,pc are {\w increasingly} incomplete  {\w in the eccentricity space} 
(see Appendix \ref{sec:selection}).

\begin{table*}
\centering
\begin{tabular}{|c| c c c c |c|} 
 \hline
 sample name & criterion & sample size & {\w mean distance} & {\w mean e-error ($\bar\epsilon_e$)} & best-fit $\sigma_e$ \\
 \hline
 \hline
 DR3 binaries & - & $362,065$ & $1240$\,pc & $0.068$ & $0.238$ \\
 `Full' & main-sequence, significance $>10$  & $147,634$ & $560$\,pc& $0.058$ & $0.240$\\
`Primordial'  &  period $P \in [10^2, 10^3]$ d  & $103,590$ & $607$\,pc & $0.065$ & $0.256$\\
 - & `Orbital' or `AstroSpectroSB1'  & $97,458$ & $618$\,pc & $0.065$ &  $0.258$\\
- & goodness-of-fit cut (see text) & $90,376$ & $629$\,pc & $0.065$ &  $0.257$\\
 `Sun-like' & primary $R_* \in [0.7,2.0]R_\odot$    & $81,107$ & $595$\,pc & $0.064$ & $0.258$ \\
`gold' &  distance $< 150$\,pc   & $3,071$ & $111$pc & $0.028$ &  $0.303$\\
  \hline
\end{tabular}
\caption{How we pare down the Gaia binary catalog to obtain various samples in this study. The right-most column lists the best-fit Rayleigh modes. {\w These values are reduced from the true one by increasing distance and therefore increasing detection bias. They are, to a lesser degree, inflated by measurement errors (see Appendix \ref{sec:error}).}}
\label{tab:sample}
\end{table*}

\begin{figure*}
    \centering
\includegraphics[width=1.0\textwidth]{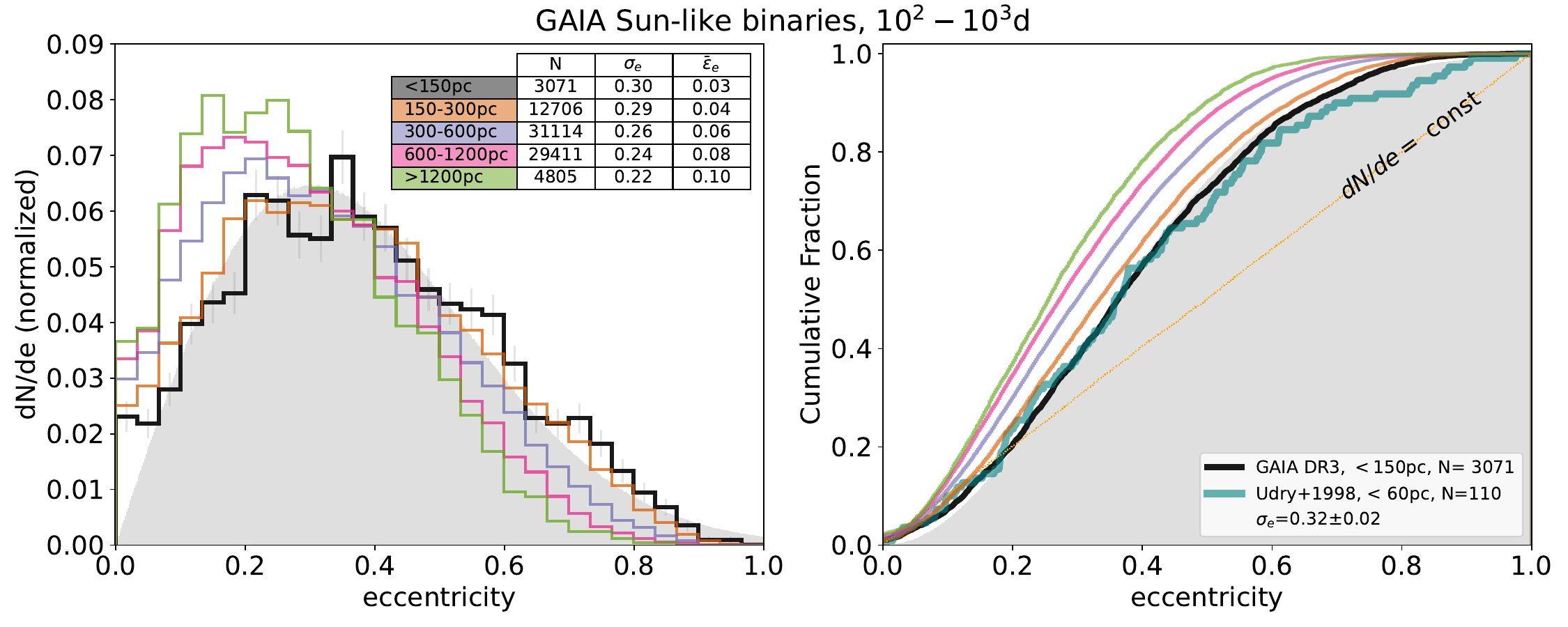}
\caption{Eccentricity distribution 
for the Gaia `Sun-like' sample (Table \ref{tab:sample}), split by distances, shown in differential (left panel) and cumulative (right panel) forms. The `gold' sample ($<150$\,pc, thick black lines) is least affected by selection 
{\w bias in eccentricity} and is well described by a Rayleigh distribution with a mode $\sigma_e=0.30$ (gray-shaded areas). E-distributions for binaries at larger distances gradually shift to the left, reflecting the increasing severity of selection bias {\w (not corrected here). They also suffer larger measurement uncertainties ($\bar\epsilon_e$, also not corrected here).} 
The right panel also presents the 
{\w $N=110$}
spectroscopic binaries within the same period range {\w from \citet{Udry1998}.} 
The Gaia gold sample is statistically consistent with this set; {\w both}
firmly
reject the `uniform' hypothesis ($dN/de \propto e^0$).
}
\label{fig:data1}
\end{figure*}

\begin{figure*}
    \centering
    \includegraphics[width=1\textwidth]{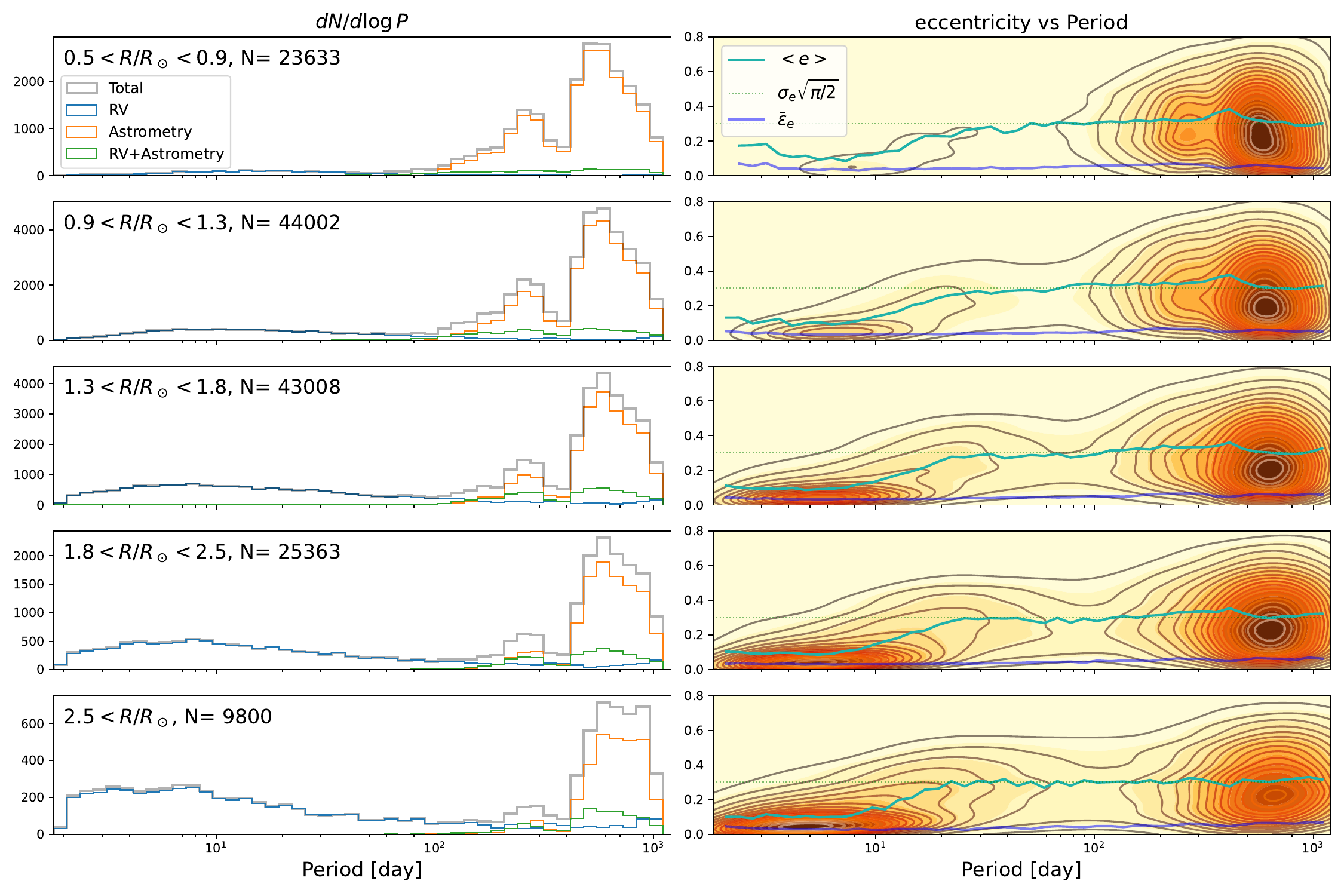}
    \caption{Properties of the `full' sample from Table \ref{tab:sample}, split by primary radii (rows) and orbital periods (x-axis). The left panels show the period distributions, indicating different detection methods. The right panels show kernel density estimation in the $e - \log P$ plane. The effects of tidal circularization are seen for binaries short-ward of $\sim 20$\,days. But beyond that, all bins reach the same mean-eccentricity ($\langle e \rangle$, green curves), corresponding to a Rayleigh mode of 
    $\sigma_e = \langle e \rangle/\sqrt{\pi/2} = 0.25$. 
    The low-lying blue curves are the mean eccentricity errors (${\bar\epsilon}_e$), with individual  uncertainties ranging typically from $0.02-0.05$. }\label{fig:mainsequence}
\end{figure*}

The eccentricity distributions of our binary samples are shown in Fig. \ref{fig:data1}. 
They are all well described by a Rayleigh distribution,\footnote{Adopting a Bayesian framework (Appendix \ref{sec:model-compare}), one can formally establish that this single Rayleigh distribution is favoured over {\w a number of} other distributions, and that its mode is strongly constrained.}
\begin{equation}
    \frac{dN}{de}  = {e\over{\sigma_e^2}} \, \exp\left({- \frac{e^2}{2\sigma_e^2}}\right)\label{eq:rayleigh}\, .
\end{equation}
For the `gold' sample, we obtain $\sigma_e = 0.303\pm 0.003$, where the uncertainty {\w only} accounts for the sample size. {\w In the mean time,  $\sigma_e$ reduces for samples at increasing distances.}

{\w In Fig. \ref{fig:data1}, we also compare}
the `gold' sample  
{\w against the sample of spectroscopic G-dwarf binaries from \citet{Udry1998}. The two are highly consistent with a p-value of $0.7$ for a two-sample KS test. 
This suggests that the eccentricity bias in the \citet{Udry1998} sample is light enough, one could have arrived at our conclusion  using just the RV sample and a quarter century earlier}.\footnote{\w However,  a comparison with the SB9 RV sample \citep{SB9} within the same period range returns a p-value of $10^{-10}$. There are more low-e binaries in the SB9 sample, which contains more red giants and likely is more strongly affected by tidal circularization.}

{\w The reduction of the Rayleigh mode at larger distances is explained in Appendix \ref{sec:selection}.}
(Figs. \ref{fig:kareem} \& \ref{fig:x_extrapolate}). This results from detection bias in Gaia \citep{elbadry2024}: more eccentric binaries are harder to detect at larger distances, because they spend more time at slowly-moving apoapsis, and their photo-centre shifts are smaller at periapsis. 
{\w Even the 'gold' sample suffers a small degree of erosion. Correcting for incompleteness using the forward-modeling pipeline of \citet{elbadry2024}, we find  that the best-fit Rayleigh mode for the 'gold' sample rises from $\sigma_e = 0.303$ to $\sigma_e = 0.336$ (Appendix~\ref{sec:selection}).}

Bearing in mind the {\w above} detection bias, we can employ the $150,000$-strong `full' {\w Gaia} sample to paint a more nuanced picture of the e-distribution. In Fig. \ref{fig:mainsequence}, we split the 
 `full' sample apart by primary mass (with primary radii from $0.5$ to $3 R_\odot$, corresponding to spectral types A-F-G-K-M) and by orbital periods (from $2$ days to $1200$d). We  adopt the mean eccentricity  $\langle e \rangle$ as a proxy for the Rayleigh mode, where
$\langle e \rangle = \sigma_e {\sqrt{\pi/2}}$ for a Rayleigh distribution.
We find that every bin, with the exception of those that are closer than $\sim 20$\,d and have therefore undergone various degrees of tidal circularization, 
exhibits a similar mean eccentricity of $\langle e \rangle\sim 0.31$, or $\sigma_e \sim 0.25$. This latter value is close to that obtained for the `Sun-like' sample (`gold' but including all distances), $\sigma_e = 0.258$, and is a result of convolution between the true mode and detection bias.  Assuming the same detection bias across all bins,\footnote{The real bias may vary from bin to bin. However, even the most biased sample in Fig. \ref{fig:data1} (those outside 1200\,pc) still returns a $\sigma_e = 0.225$, or a value lower by $\sim 25\%$ from $\sigma_e = 0.30$. Such a relative immunity from selection bias is offered by the fact that a Rayleigh distribution contains mostly low-e systems that are less vulnerable to incompleteness.} this exercise suggest that the intrinsic Raleigh mode is invariant, across a wide range of primary spectral types, and over a large span in orbital periods. Such an invariance is remarkable and points to a universal process at work. 

\begin{figure*}
    \centering
\includegraphics[width=0.98\textwidth]{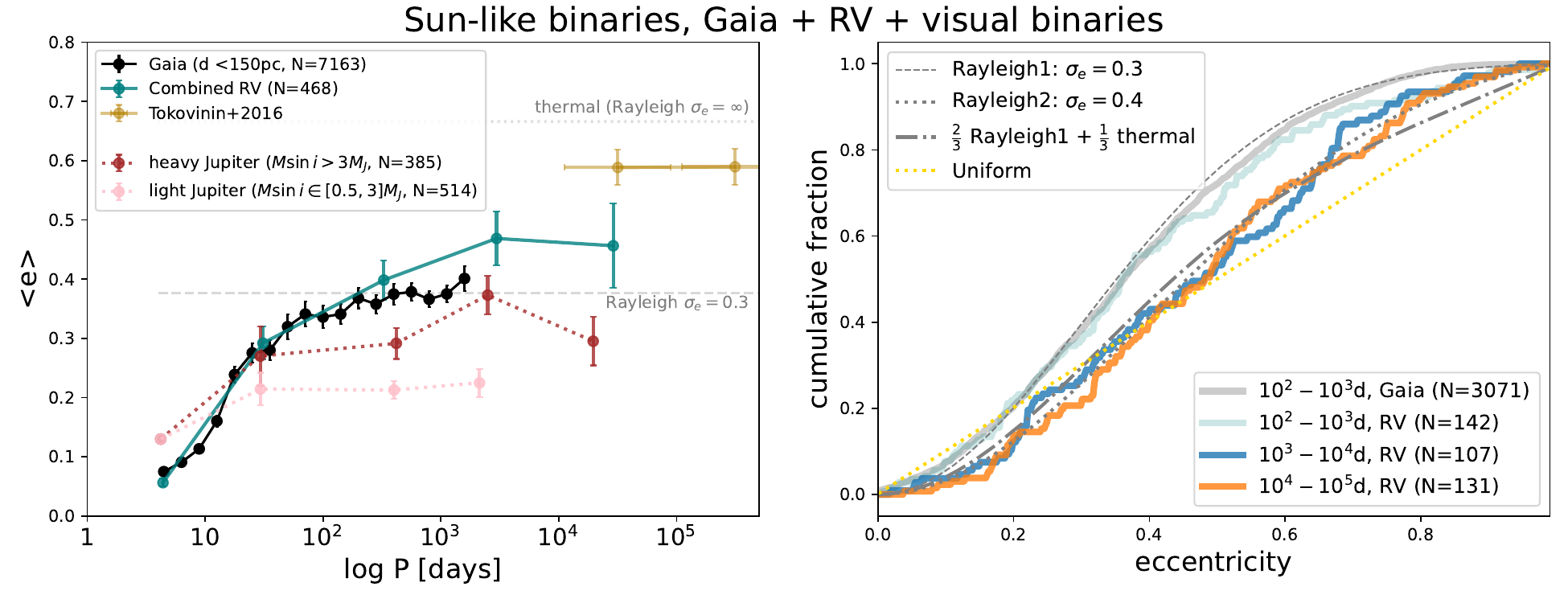}
    \caption{{\w Exploring  the eccentricity distribution beyond Gaia binaries. 
    The left  panel plots the mean eccentricity as a function of orbital periods, for stellar and planetary companions. RV data are combined from \citet{Udry1998,Raghavan2010,Griffin2012}; visual binary data are from  \citet{Tokovinin2016}; planet data are taken from exoplanetarchive.ipac.caltech.edu and are split into light and heavy Jovian companions.
  Stellar binaries beyond $10^3$ days exhibit rising mean eccentricities. The right panel compares these long period binaries  against some models.}
}\label{fig:all_rv}
\end{figure*}

{\w We also investigate whether the Rayleigh mode depends on the mass ratio within a binary. We fail to reach a definitive conclusion based on current Gaia data (Appendix \ref{sec:massratio}, Fig. \ref{fig:massratio}).}
{\w On the other hand, a piece of oblique evidence suggests that it does not:  twin binaries are abundant among spectroscopic binaries \citep[see, e.g.][]{Tokovinin2000} but are largely absent in Gaia as their photo-centres hardly shift; so the agreement between RV and Gaia samples (Fig. \ref{fig:data1}) suggests that the eccentricity distribution is largely independent of the binary mass ratio.}
{\w Such an independence may even extend to the planetary regime. Fig.\,\ref{fig:all_rv} suggests that the eccentricities of massive Jovian planets are similar to those of stellar companions. }


Lastly, we wish to gain some insights as to {\w the outer reach of} the Rayleigh distribution. \citet{Tokovinin2020,Hwang2022}
show that binaries outside 100AU ($\sim 3 \times  10^5$ d) are thermally distributed in their eccentricities ($dN/de \propto e$). 
What about the binaries between a few AU ($10^3$ days) and 100AU? {\w There can be at least three possibilities: the Rayleigh distribution may persist but with a mode that increases with period, smoothly connecting up to the thermal distribution (the ``rising Rayleigh'' model);\footnote{Formally, the thermal distribution, $dN/de \propto e$, is also a Rayleigh distribution but with $\sigma_e \rightarrow \infty$.} alternatively, the Rayleigh mode may remain constant but the Rayleigh distribution is gradually replaced by a thermal one as the period increases (the ``mixture'' model); lastly, the in-between group can take the ``uniform'' form ($dN/de =$ const).}

{\w Impressed by the constant Rayleigh mode we observe in the Gaia sample (Fig. \ref{fig:mainsequence}), we hypothesize that the ``mixture'' model is correct.
We are also} prejudiced by the following theoretical argument.
Close binaries are thought to extend to tens of AU, likely determined by the size of massive disks \citep{Tobin2020} and by the scale at which gravitational collapse occurs \citep{rafikov05,matznerlevin05,kratter10}. {\w If all close binaries experience the same eccentricity excitation, then}
the same Rayleigh distribution should be observable out to tens of AU. 

{\w We evaluate this hypothesis by examining available data. Binaries from $10^3$ to $\sim 10^5$ days are probed by RV and other techniques. We combine the RV samples from \citet{Udry1998,Griffin2012,Raghavan2010} to enhance the long-period sensitivity, and adopt the result for visual binaries from  \citet{Tokovinin2016}. As Fig. \ref{fig:all_rv} shows, the mean eccentricities clearly rise beyond $10^3$ days \citep{Tokovinin2016}. Unfortunately, the rising Rayleigh and the mixture model look very similar. Both are currently  permitted by data (right panel of Fig. \ref{fig:all_rv}). The uniform model, on the other hand,  is disfavored.}
{\w In addition,} while \citet{Tokovinin2020,Hwang2021} showed that a `uniform' model can explain  the instantaneous radial-velocity vectors for Gaia 
resolved binaries  \citep[$10-100$AU:][]{Elbadry2021}, Fig. \ref{fig:mixed} argues that 
both the rising Rayleigh and the mixture models can also do the job. 

{\w In conclusion, for binaries from a few to 100AU, while the `uniform' model may be discarded, it will take a much larger sample\footnote{\w We estimate that an un-biased sample of $N\geq 2000$ is needed.} to distinguish the `rising Rayleigh' model from the `mixture' model.} This may be possible using the data promised by Gaia DR4{\w /DR5.}




\section{Possible Origin{\w s}}
\label{sec:theory}

The eccentricities of AU-scale binaries are likely primordial, not affected by their birth clusters \citep[Appendix \ref{sec:unimportance}, also][]{Parker2009,Spurzem2009}, even less so by passing-by stars {\w after the cluster dispersal}. Our finding of a Rayleigh distribution that is invariant points to a universal process at birth.
{\w We briefly describe two possibilities: one is excitation by the circumbinary disks, and one is by ejecting brown dwarfs. The latter is a new proposal.}

\subsection{\w Circumbinary Disks}
\label{subsec:disk}

{\w Circumbinary disks, present around young binaries, may excite their eccentricities.  This is an insight gained from numerical simulations  \citep[for a review, see][]{Lai2023}. These simulations also found that there exist `attractor states', where the binary eccentricity reaches an equilibrium at around  $e \sim 0.3 - 0.5$ \citep{Munoz2019,Orazio2021,Zrake2021,Siwek2023,Valli2024}. The exact value depends on parameters such as binary mass ratio, disk viscosity and thermodynamics. It is argued that even a disk with only a few percent of the stellar masses suffices for the task \citep{Dittmann2024,Valli2024}. 
}

{\w While we have advocated for a Rayleigh fit for its simplicity (one parameter fit), the Gaia binary data can also be described by a two-parameter Gaussian distribution 
\begin{equation}
    \frac{dN}{de}  = \exp\left[{- \frac{(e-e_0)^2}{2\sigma^2}}\right]\, ,
\label{eq:normal}
\end{equation}
with $e_0 = 0.38$ and $\sigma = 0.20$ for the `gold' group. In this case, one can interpret $e_0$ as the most probable value from circumbinary action, and $\sigma$ the dispersion that arises from differences in physical parameters. }


{\w Interestingly, the theory community have long predicted eccentricity excitation by circumbinary disks  and have been puzzled by the lack of signature in the observed binaries. Our result seems to validate this theory.}

{\w Before a victory is declared, however, we note that such a theory is still in its infancy. The origin of the `attractor state' has not been elucidated. The simulations are still too simplified, for instance, in their treatments of disk thermodynamics \citep{Wang2023}. 
It is unclear why the observed eccentricity distribution is independent of binary period and stellar mass (Fig. \ref{fig:mainsequence}). 
Lastly, current models suggest that the equilibrium eccentricity rises with binary mass ratio \citep{Siwek2023,Valli2024}, at odds with what we infer above. 

These disagreements motivate us to expore an alternative scenario below.}


%
%


\subsection{\w Scattering Brown Dwarfs}
\label{subsec:scatter}

{\w This scenario is suggested by t}he Rayleigh form itself. A Rayleigh distribution is simply a Gaussian in 2-D: the two components of the eccentricity vector are each normally distributed around zero with the same mode.\footnote{In other words, a random Gaussian distribution in the Cartesian velocities. For a more general form of a triaxial Gaussian distribution, see \citet{Greenzweig1992}.
} Such a situation naturally arises from weak random scatterings, as has been observed in numerical experiments of planetesimal scatterings \citep{Greenzweig1992,Ida1992,Tremaine2015}.  
In fact, gravitational scatterings have been invoked to explain  the  e-distribution of asteroids \citep{Malhotra2017}, and exo-planets  \citep{Zhou2007,Juric2008,Ford2008}.
This line of thought stimulates our investigation of the following scenario.

\begin{figure*}
    \centering
    \includegraphics[width=0.48\textwidth]{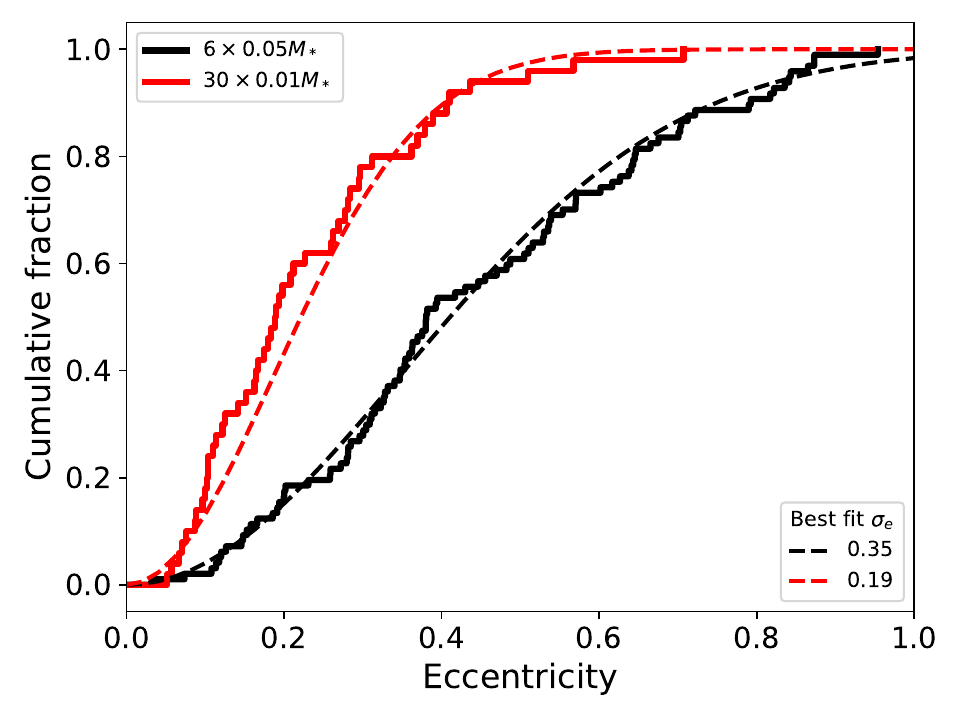}
    \includegraphics[width=0.48\textwidth]{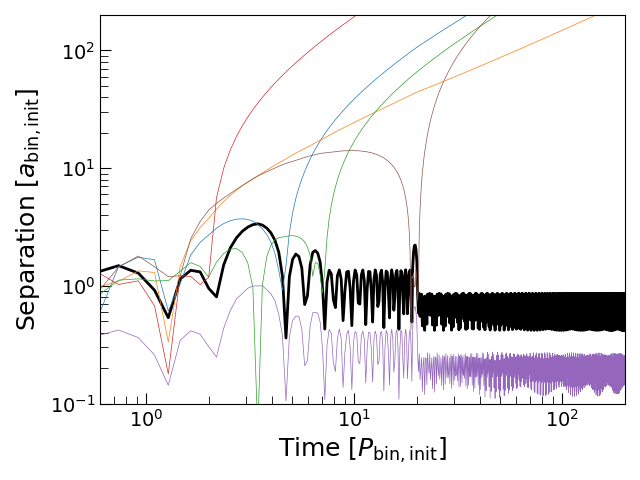}
    \caption{Results of scattering experiments.  We populate binary systems (primary $0.7 M_\odot$, secondary $M_*=  0.3 M_\odot$) with brown dwarfs that have the same total mass but different individual masses. The final binary e-distributions are Rayleigh in form (left panel, 100 cases each), with modes that depend only on the individual brown dwarf mass (eq. \ref{eq:equi2}). The right panel shows an example case with the orbits of the brown dwarfs as colored curve, and the secondary as heavy black curve. The vertical axis is the distance to the binary centre of mass. All brown dwarfs are promptly ejected except for one that becomes bound to the primary.
}\label{fig:scattering}
\end{figure*}

Consider, at birth, the presence of one or more low-mass bodies (`brown dwarfs') in the binary system. Unless sheltered dynamically, they should be quickly ejected by the binary through close encounters. Such encounters tend to establish equi-partition of 
epicyclic energy among the bodies \citep{Ida1992b},
\begin{equation}
    e_*^2 M_* \simeq e_{\rm BD}^2 M_{\rm BD}\, ,
    \label{eq:equipartition}
\end{equation}
where $M_*$
is the mass of the secondary. Setting $e_{\rm BD} \sim 1$ for ejection, we obtain, 
\begin{equation}
    e_* \simeq \left(\frac{M_{\rm BD}}{M_*}\right)^{1/2} 
    \simeq 0.30 
    \left(\frac{M_{\rm BD}}{0.03M_\odot}\right)^{1/2}\, 
    \left(\frac{M_*}{0.3M_\odot}\right)^{-1/2} \, ,
    \label{eq:equi2}
\end{equation}
This expression does not depend on period, because 
the scattering dynamics is scale-free \citep[ejection velocity is related to  the local Keplerian velocity; also see Fig. 8 of][]{Juric2008}. Moreover, if there is (even a weak) correlation between $M_{\rm BD}$ and $M_*$,  this expression will also be insensitive to the stellar masses. These will then explain the {\w universal Rayleigh mode} we find  in Gaia binaries.

We carry out numerical experiments (details in Appendix \ref{sec:scatterings}) and present the results in Fig. \ref{fig:scattering}. We endow each binary system with a number of low-mass siblings. As the brown dwarfs are promptly cleared away, the binary eccentricities acquire a Rayleigh distribution, with the mode  described roughly by eq. (\ref{eq:equipartition}). To achieve the observed value of $\sigma_e = 0.30$, we require $M_{\rm bd} \sim  M_*/10$. We find that the outcome is not sensitive to the total number of brown dwarfs -- in fact,  {\w even one} brown dwarf per system suffices  \cite[also see][]{Ford2008}.

Could Nature provide such a set-up consistently?
In the scenario of disk fragmentation,
it is  plausible that multiple low-mass objects would form alongside the dominant binary. At gravitational fragmentation, the characteristic mass is $\sim 10^{-2}M_\odot$
\cite[e.g.,][]{rafikov05,matznerlevin05,Xu2024}. This is close to the above brown dwarf mass. 
Subsequent nonlinear evolution is currently unclear \citep{Goodman2004,Levin2007}. The fragments can accrete from the disk and/or merge with each other. They can also migrate due to disk torque. The prevalence of close binaries in nature suggests that, in many cases, some of these seeds can grow to reach the isolation mass ($\sim M_\odot$). At the same time,  other small seeds may persist or may be continuously produced, as in the forming triple system observed by \citet{tobin16triple}. 
If so, disk fragmentation sets the stage for later dynamical scattering. 
Further investigations may reveal if $M_{\rm bd}/M_*$ does remain nearly constant in all disks. 

{\w Previously, 
\citet{Reipurth2001} and \citet{Kroupa2003} have proposed that brown dwarfs are ejected embryos from their natal systems. Our hypothesis is similar in nature to theirs.} 
If all close binaries have ejected 
{\w about} one brown dwarf, this can explain all sub-stellar objects detected by imaging in young clusters and by microlensing surveys (Appendix \ref{sec:enoughbd}). 
{\w Moreover, one expects these brown dwarfs to acquire velocity dispersions of order the orbital escape velocity.\footnote{\w For a 50\,au binary (peak of the binary) with $M_1 = 0.3 M_\odot$ (peak of the initial mass function), this is $\sim 2\km/\s$.}   If this is larger than the dispersion velocity in a young cluster, one expects that brown dwarfs should preferentially lie at the outskirts of these clusters, and that they have larger proper motions. Currently, there are conflicting observational evidences \citep[e.g.][]{Kumar2007,Scholz2012,Downes2014,Panwar2024}, and a systematic study is required.}

\subsection{\w Planets and Triples}

{\w As both theories (disk and scattering) suffer from insufficiencies, we look to data for clues. Here, we examine two populations:  Jovian planets, and compact hierarchical triples.}

{\w Fig. \ref{fig:all_rv} shows that heavy Jovian planets ($M\sin i > 3 M_J$) in the period range ($10^2-10^3$ days) exhibit a similar eccentricity distribution as Gaia binaries. In comparison, lighter planets are more circular. Both disk and scattering may potentially explain these facts. Heavy planets can grow in eccentricity by interacting with the disks \citep[see, e.g.][]{Papaloizou2001}; they could also be excited by destabilizing and scattering their siblings \citep[see, e.g.,][]{Ford2008,Juric2008}. The latter requires the ejected bodies to be dominated in mass by those that are  $\sim 1/10$ of the Jovians. It is unclear if this is possible. Similarly, the disk theory also needs to explain why systems of such low mass ratios are excited to the same eccentricities as stellar binaries. }

{\w Compact hierarchical triples are stellar systems where the outer orbits are smaller than a few AU \citep[see review by][]{Tokovinin2021}.
The eccentricity distribution of their outer orbits is also similar to that of the Gaia binaries considered here 
\citep{Borkovits2016,Cza2023,Moharana2024}\footnote{Their inner orbits are commonly short and have likely experienced tidal damping.} This suggests that the presence of a tight inner binary does not affect the outer eccentricity.  
Moreover,  \citet{Borkovits2016,Tokovinin2017} demonstrated a strong degree of alignment between the inner and the outer orbits in these systems, with a dispersion of order $20\deg$.
This is, again, naturally accommodated by both the disk and the scattering theory (see Appendix \ref{sec:scatterings}).

In conclusion, we remain in limbo.  
}

\section{Conclusions}\label{sec:conclusions}

The Gaia mission greatly expands our sphere of vision. The number of AU-scale binaries, for which eccentricity information can be reliably extracted, rises 
{\w from thousands} to of order $10^5$.  Among these, we select a largely un-biased sample of $\sim 3000$ systems to deduce the underlying eccentricity distribution.
A simple and elegant Rayleigh distribution emerges, with a mode of $\sim 0.3$. The value of the mode appears invariant with respect to stellar mass and orbital period, but a more definitive conclusion will require careful study of the selection bias.

Such a distribution is almost certainly primordial in origin. Its (apparent) invariance points to a universal process. And the Rayleigh form itself suggests an origin in weak scatterings. 

{\w Our numerical experiments show that, in order} to reproduce the observed Rayleigh mode,  the stellar binaries need to eject brown dwarfs with masses  $\sim 1/10$ as heavy.
It is not known why this must be so. But if true, such ejections can account for almost all free-floating 
{\w brown dwarfs}. The kinematics  {\w of these bodies should} bear imprints of the ejection process.
%

{\w An alternative theoretical explanation also exists. Binaries may spontaneously grow in eccentricity when interacting with circumbinary disks. While this behavior has been predicted by numerical simulations, theoretical understanding is lacking and current predictions do not readily agree with data. 
We point out that stars with planetary companions, and stars in triple systems, may provide extra constraints on the excitation.
}


{\w The radial extent of the Rayleigh distribution could also inform about its origin. Available data suggest that the mean eccentricity rises with binary period. But it is unclear if the Rayleigh mode itself rises beyond $10^3$ days.} {\w We hope that Gaia 
DR5, with a baseline of 11 yrs, } will provide  an answer  to this question.


With this new discovery, 
the eccentricity distribution now becomes a new marker for the process of disk fragmentation, joining rank with other measurables like period, metallicity, mass ratio, and twin-fraction. It can be used to probe many interesting questions. For instance, do binaries in extreme environments (e.g., globular clusters, nuclear star clusters) form in disks?  Does the Rayleigh mode vary with environmental factors? 

AU-scale binaries are important drivers for binary stellar evolution -- given the AU-sizes of giant stars, these include most of the binaries that are destined to interact during their lifetimes,  through tides, mass transfer and common envelope. 
The Rayleigh distribution, as opposed to the older uniform distribution, leads to fewer binary mergers. This new distribution should be adopted in synthetic studies of binary evolution. 




%



\bigskip

{\w We thank an anonymous reviewer for penetrating comments that lead to a much expanded discussion, and Ayush Moharana for bringing the compact hierarchical triples to our attention.} We thank NSERC for research funding, and the Gaia collaboration for a marvelous gold mine. 




\bibliography{binary}

\begin{appendix}

\section{GAIA binary extraction}\label{sec:A1}

We construct a binary sample that is little affected by either detection bias or evolutionary changes.

Starting from the Gaia non-single-star catalog \citep{Arenou2023}, we apply a number of cuts consecutively.  These steps and their resulting sample sizes are listed in Table \ref{tab:sample}. Our final sample (named the `gold' sample) is 
homogeneous in primary properties (Sun-like dwarfs), and avoid any significant selection bias.

Here are some explanations. A fraction of the Gaia catalogue stars have determined astrophysical parameters. Among these, we retain only systems with main-sequence primaries, defined as
\begin{equation}
M_G  > {\rm Max}\{4.0, \, \left[(B_p-R_p)_{\rm de-red} - 1\right]^7\}\,
\label{eq:BPRP}
\end{equation}
where $M_G$ is the absolute g-band magnitude and $(B_p-R_p)_{\rm de-red}$ is the de-reddened color. 
We also remove those with `significance' $<10$. Aside from producing a cleaner sample, this automatically rejects any binaries detected as `Eclipsing Binaries'. We exclude these binaries because many of them have eccentricities artificially set to zero. 

We also select binaries with periods from $10^2-10^3$days. The former is set to avoid pollution from tidal circularization, while the latter is due to the finite coverage of Gaia DR3. To meaningfully compare against the detection completeness from mock pipelines \citep{elbadry2024}, we proceed to retain only binaries characterized by astrometry (including `Orbital' and `AstroSpectroSB1' types). Among these, we require a further quality cut: `goodness-of-fit' $< 5$  if `phot-g-mean-mag $> 13$, or  $< 10$ if otherwise.\footnote{\w This quantity indicates how compatible the Gaia data is with a single star solution.} This filters out bad solutions, the threshold for which depends on brightness. A system is considered `Sun-like' if the primary radius is within $[0.7,2.0]R_\odot$.  

And lastly, for our `gold' sample, we set a maximum distance of $150$\,pc. This last cut leaves us with $\sim 3000$ systems, a tiny fraction of the original data($\leq 1\%$). About $\sim 45\%$ of the `gold' sample have only astrometric orbits (`Orbital'), while the rest are additionally characterized by radial velocity (`AstroSpectroSB1'). The latter group tend to be brighter.

Ideally, we would also like to remove systems with white-dwarf secondaries. Their eccentricities have likely been strongly suppressed during the giant phase. However, doing this thoroughly is difficult at the moment \citep{Shahaf2023}. Fortunately, such binaries mostly contribute at the longest periods. Results in Fig. \ref{fig:mainsequence} show that their impact is likely minor. 

Regardless of the cut, all samples exhibit e-distributions that have a  Rayleigh shape. Their Rayleigh modes, however, differ (Table \ref{tab:sample}).
Most cuts return a lower mode  ($\sigma_e \sim 0.25$), except for the `gold' sample ($\sigma_e \sim 0.30$).\footnote{Within the `gold' group, the Rayleigh mode does not vary with brightness, nor with the detection method.}
Such a difference, we argue (Fig. \ref{fig:mainsequence}), reflects not intrinsic variation, but impact of the selection bias (Appendix \ref{sec:selection}). 



\section{Characterizing the Detection Bias}
\label{sec:selection}

\begin{figure*}
\centering
\includegraphics[width=0.95\textwidth]{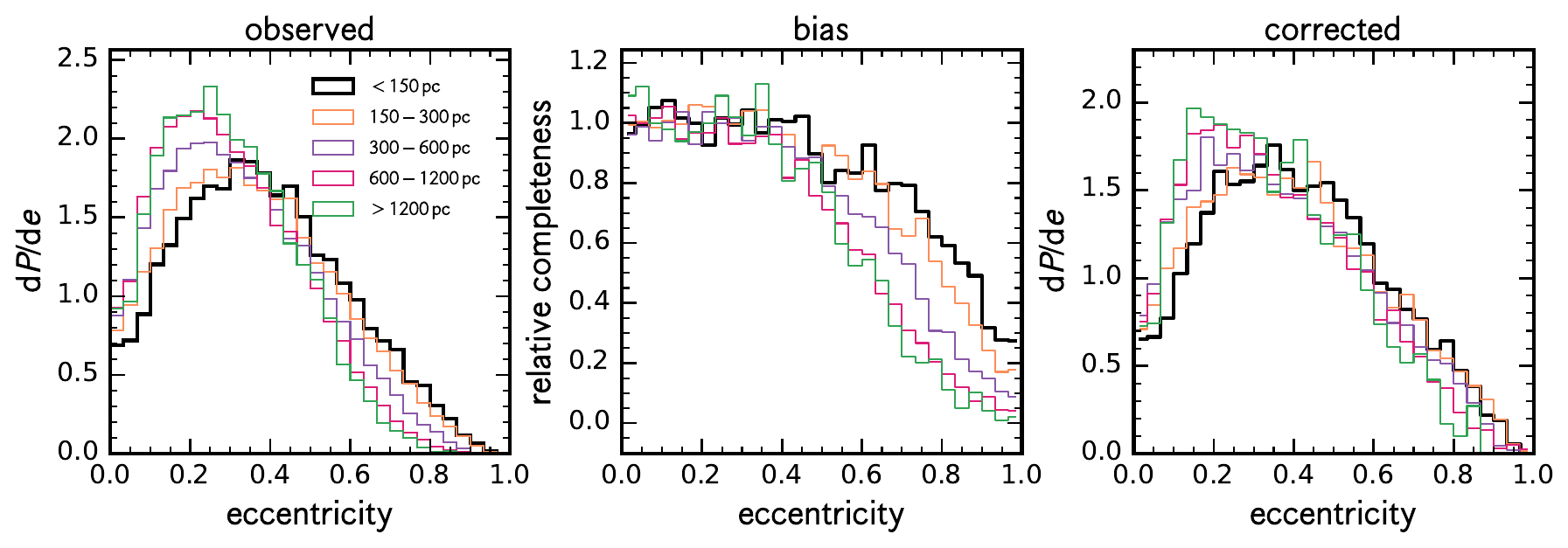}
\caption{Correction for detection bias. The left panel repeats that in Fig. \ref{fig:data1}. The middle panel shows the completeness as a function of binary eccentricity, obtained from the mock astrometry pipeline \citep{elbadry2024}.  The right panel shows the corrected values. Correction brings the different distributions into better agreements with each other, though some differences still exist. {\w Measurement errors, not included here, may help resolve some of these differences.}
}
\label{fig:kareem}
\end{figure*}

{\it Gaia} is, generically, less sensitive to high-eccentricity orbits. High-eccentricity binaries spend more time near apoapse and are thus less likely to have their orbits well sampled by {\it Gaia} observations, which occur at quasi-random times. This results in orbits that are on average less well constrained at high eccentricity, and less likely to pass the stringent quality cuts imposed on the orbital solutions published in {\it Gaia} DR3. Such a bias very likely affects both astrometric and RV orbits, but it has thus far been quantitatively modeled only for astrometric orbits, and for this reason we focus on astrometric orbits in this work. 

We use the forward-model of {\it Gaia} astrometric orbit catalogs described by \cite{elbadry2024} to quantify the eccentricity bias in our sample. The model produces realistic epoch astrometry for each simulated binary and fits it using the same cascade of astrometric models used in producing the DR3 binary catalogs \cite[see][]{Halbwachs2023}. Following \cite{elbadry2024}, we generate a realistic population of binaries within 2 kpc of the Sun, mock-observe them, and produce a mock catalog. Unlike \cite{elbadry2024}, who assumed an eccentricity distribution following \cite{Moe2017}, we  adopt a uniform eccentricity distribution, 
which allows us to trace {\it Gaia's} eccentricity bias. Other properties of the simulated binaries (e.g. masses, ages, evolutionary states, orbital periods, 3D locations in the Galaxy) are chosen as described by \cite{elbadry2024} and are reasonably good approximations of reality. To represent the bias specific to the observational sample analyzed here, we exclude binaries containing red giants and require astrometric significance $> 10$ in addition to the quality cuts imposed on the solutions published in DR3. 

The results of these simulations are shown in the middle panel of Figure \ref{fig:kareem}. As with the observed data, we show individual distance bins separately. We normalize the distributions in each eccentricity bin such that the average relative completeness at $0 < e < 0.3$ is unity. A bias against high eccentricities exists in all distance bins, but it is more severe at large distances, where orbits have smaller angular size at fixed period and thus lower astrometric SNR. This trend mirrors what is found in the observed sample. The simulations suggest that {\it Gaia} sample with $d < 150$\,pc is largely unbiased at $e < 0.5$, but a bias against high eccentricities is present at higher eccentricities. At $e>0.9$, the completeness is $\approx 3$ times lower than at $e < 0.5$. 

The right panel of Figure~\ref{fig:kareem} shows the normalized eccentricity distribution of the observed samples after correcting for incompleteness. The corrected eccentricity distributions in different distance bins are in better agreement with one another, although some trend of lower eccentricity at larger distances is still present. This could reflect imperfections in the forward-model, although it could also result from a mass-dependent eccentricity distribution, since more distant binaries are on average brighter and more massive {\w (but see Appendix \ref{sec:massratio}).} {\w For the 'gold' sample, we find that these corrections bring the best-fit Rayleigh mode from $\sigma_e = 0.303$ to $\sigma_e = 0.336$.}

\begin{figure*}
\centering
\includegraphics[width=0.95\textwidth]{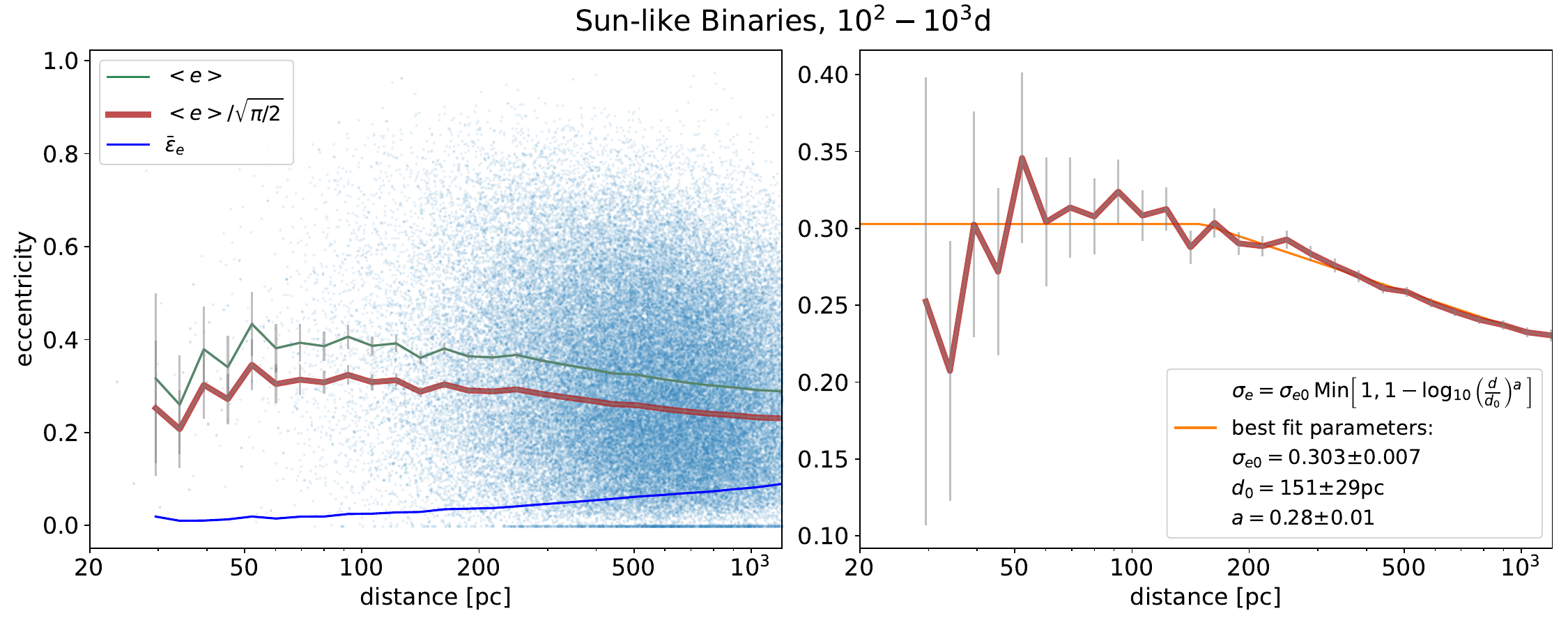}
\caption{We use the `Sun-like' sample in Table \ref{tab:sample} to explore the effects of detection bias. The left panel shows that the mean eccentricity ($\langle e \rangle$) gradually declines with distance. The brown line is $\langle e \rangle/\sqrt{\pi/2}$, a proxy for the Rayleigh mode. Using equation \ref{eq:decline} to describe how distance affects the mode, we find an an intrinsic Rayleigh mode of $\sigma_{e0} \approx 0.30$, and that systems inward of $d_0 \sim 150$\,pc should be free of detection bias. {\w The blue curve on the left ($\bar\epsilon_e$) denotes the mean measurement error in eccentricity.}}
\label{fig:x_extrapolate}
\end{figure*}

We also use the observed sample itself to gauge how much this bias affects the measurement for the Rayleigh mode. This is a self-calibration method and is  independent of the above mock pipeline.

In Fig. \ref{fig:x_extrapolate}, we present the eccentricities and their mean values ($\langle e \rangle$) as functions of distance from Earth. We use $\langle e \rangle/\sqrt{\pi/2}$ as a proxy for the Rayleigh mode $\sigma_e$ and fit the following data-inspired form,
\begin{equation}
\sigma_e (d) = \sigma_{e0} \, {\rm Min}\left[1, 1- \log_{\rm 10}\left(\frac{d}{d_0}\right)^a\right]\,.
    \label{eq:decline}
\end{equation}
where $\sigma_{e0}$ is the intrinsic Rayleigh mode. Such a form asserts that, 
for systems closer than $d_0$, $\sigma_e = \sigma_{e0}$, i.e., there is little bias for the bulk of the Rayleigh distribution (but there can still be bias at high eccentricities).
Such a form is reasonable for a Rayleigh distribution (which concentrates at low-e), but is less so for a, e.g., power-law distribution.

The observed data yields the following best-fit parameters: $\sigma_{e0} = 0.301\pm 0.006$, $d_0 = 147 \pm 27$\,pc, and $a=0.29\pm 0.01$. This validates our main results, that the intrinsic Rayleigh scale is $\sigma_e \sim 0.3$, as measured from the `gold' sample. This does not mean, however, that the `gold' sample is complete to $d_0  \sim 150$\,pc -- it is not. It is missing binaries of high eccentricities (see Fig. \ref{fig:kareem}), as well as binaries of comparable brightnesses, or those with very low-mass secondaries. {\w But these deficiencies do not strongly affect the Rayleigh mode.}

{\w We notice a strange population of very circular binaries in Fig. \ref{fig:x_extrapolate}.  They account for $\sim 3\%$ of the total system and have reported eccentricities that fall much below the measurement errors. This is also seen in the excess in low-eccentricity bins in Fig. \ref{fig:data1}. They likely arise from issues in the Gaia pipeline. Removing this population entirely does not change the Rayleigh mode we inferred. 
}

\section{\w Eccentricity Errors}
\label{sec:error}

{\w We now turn to consider the impact of eccentricity errors. Fig. \ref{fig:x_extrapolate}  presents the mean measurement errors ($\bar \epsilon_e$) as reported by {\it Gaia}. They rise with distance, with $\bar\epsilon_e \sim 0.02$ in the nearest systems to $\sim 0.10$  at $1.5$\,kpc. Eccentricity errors enlarge the Rayleigh mode and we model it as the following.
Consider one component of the eccentricity vector.  Let $e_x$ be the original value with a standard derivation of $\sigma_e$, noise $\delta _e$ be normally distributed around zero with a standard deviation $\epsilon$ (with $\epsilon = \bar\epsilon_e/\sqrt{\pi/2}$). The measured value $e'_x$ should be distributed as
\begin{eqnarray}
& & P(e'_x) =  \int_{-\infty}^{+\infty} P(e_x) P(\delta_e) d\delta_e\, 
\nonumber \\
& & \propto \int_{-\infty}^{+\infty} \exp\left[- \frac{{(e'_x - \delta e)}^2}{2\sigma_e^2}\right]\, \exp\left[- \frac{\delta_e^2}{2\epsilon^2}\right]\, d \delta_e\nonumber \\
& & \propto  
\exp\left[ - {{ab}\over{a+b}} {e'_x}^2\right]\, 
\int_{-\infty}^{+\infty}
\exp\left[ - (a+b)\left(\delta_e - {{a e_x'}\over{a+b}}\right)^2
\right] \, d\delta_e\nonumber \\
& & \propto  
\exp\left[ - \frac{{e'_x}^2}{2 (\sigma_e^2 + \epsilon^2)}\right]\, , 
   \label{eq:convolution}
\end{eqnarray}
where $1/a = 2 \sigma_e^2$ and $1/b = 2 \epsilon^2$. So the measured Rayleigh mode for $e'$ is now larger and is $\sqrt{\sigma_e^2 + \epsilon^2}$, as is dictated by the Central limit theorem. For the above reported measurement errors, this change is relatively minor, in comparison to those caused by selection effects.
}

\section{Dependence on Mass Ratio}
\label{sec:massratio}

{\w The Gaia NSS catalogue reports masses for the primary stars, based on isochrone fitting.\footnote{\w We adopt the IsocLum mass from \citet{Arenou2023}. There is a strange cluster at mass of $1.07214M_\odot$. We replace these with their Flame masses.}
For a subset of the main-sequence binaries, the Gaia pipeline also returns mass estimates for the secondaries. It does so by assuming that the components satisfy the main-sequence mass-luminosity relation. }
{ Using 
{\w this subset,}
\footnote{We further restrict ourselves to those with `fluxratio' $> 10^{-3}$ in order to remove any systems with white dwarf companions. This is not important for the `gold' sample but does affect the samples at larger distances.} we analyze how the eccentricities depend on mass ratio (Fig. \ref{fig:massratio}).
%
We split the `Sun-like' sample by the same distance division as before, and present results for the three closest group (with enough cases).
We find that the binary counts drop off  steeply towards equal-mass, in  contrast to that found in the spectroscopic sample by \citet{Moe2017}. This is explained, at least partly, by the fact that an equal-brightness binary exhibits zero astrometric signal. 
We use mean eccentricity as a proxy for the Rayleigh mode and find that it 
{\w appears to drop off towards equal mass in the `gold' sample, but are largely flat for the other two samples we tested.} 
{\w This drop-off} occurs at the same mass ratio as the number drop-off, suggesting a common origin in detection bias. 
{\w This leads us to conclude that we cannot at the moment extract a definitive answer from the Gaia data. }


\begin{figure*}
\centering
\includegraphics[width=0.95\textwidth]{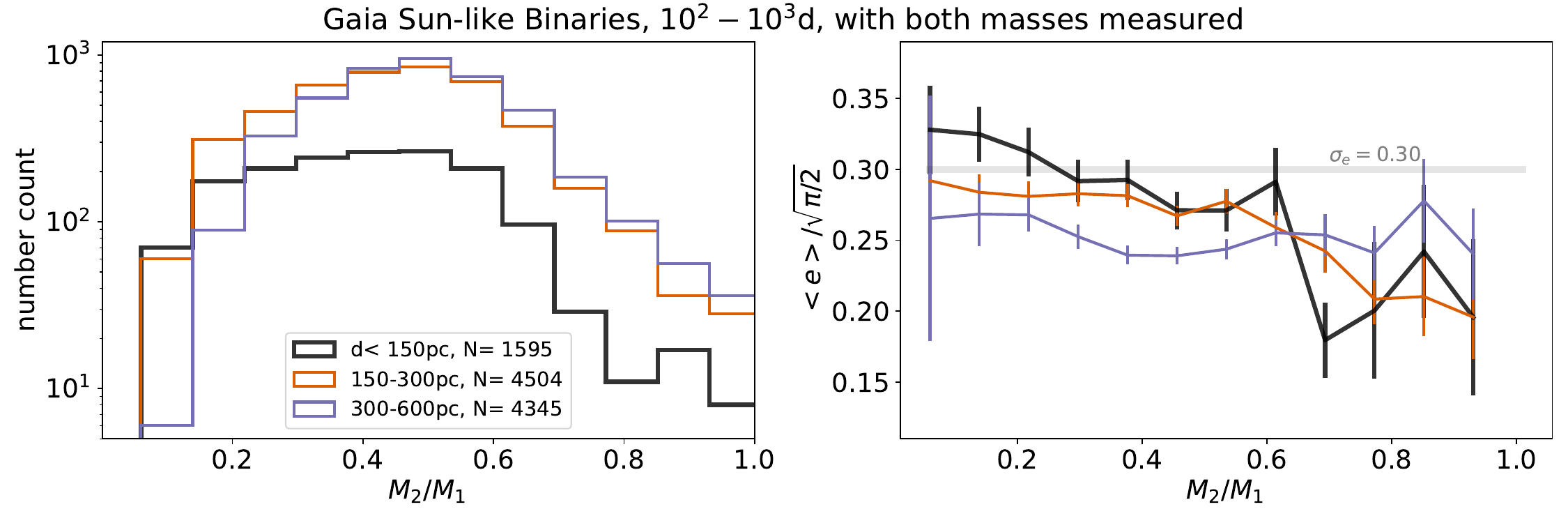}
\caption{A possible dependence of the Rayleigh mode on binary mass ratios is seen, in the Gaia `Sun-like' sample. The left panel shows the number of systems with reported mass ratios, divided by distances. The drop-off 
{\w beyond mass ratio $0.6$}
may be explained by astrometric bias {\w against detecting equal brightness pairs}.
The right panel shows the proxy for the Rayleigh mode, $\langle e \rangle/\sqrt{\pi/2}$. A possible reduction towards equal-mass is seen, but we remain suspicious.
}
\label{fig:massratio}
\end{figure*}



\section{Binaries at Longer Periods?}
\label{sec:longperiod}

{\w Here, we re-examine the resolved binaries from Gaia \citep{Elbadry2021}}. \citet{Tokovinin2020,Hwang2022} established, statistically, the eccentricity distribution by measuring the instantaneous velocity-position ($v-r$) angles that are projected on the sky 
Results from \citet{Hwang2022} are shown in Fig. \ref{fig:mixed} for binaries with projected separations that fall within $10-1000$AU. 
In the absence of a better model, they parameterized the eccentricity distribution as a power-law, $dN/de \propto e^\eta$ and found that the value of $\eta$ rises from 0 (`uniform') at $\sim 50$ au,  to $1$ (`thermal') at $\sim 500$au,  and $>1$ (`super-thermal') beyond. Alternatively, as we show in Fig. \ref{fig:mixed}, their so-called `uniform' distribution can easily be swapped for a mixture model with roughly comparable fractions in the thermal distribution and the Rayleigh distribution. 
In fact, the apparent change in $\eta$ with period \citep{Hwang2022} may simply reflect the decreasing fraction of the Rayleigh with increasing period.

\begin{figure}
\centering
\includegraphics[width=0.45\textwidth]{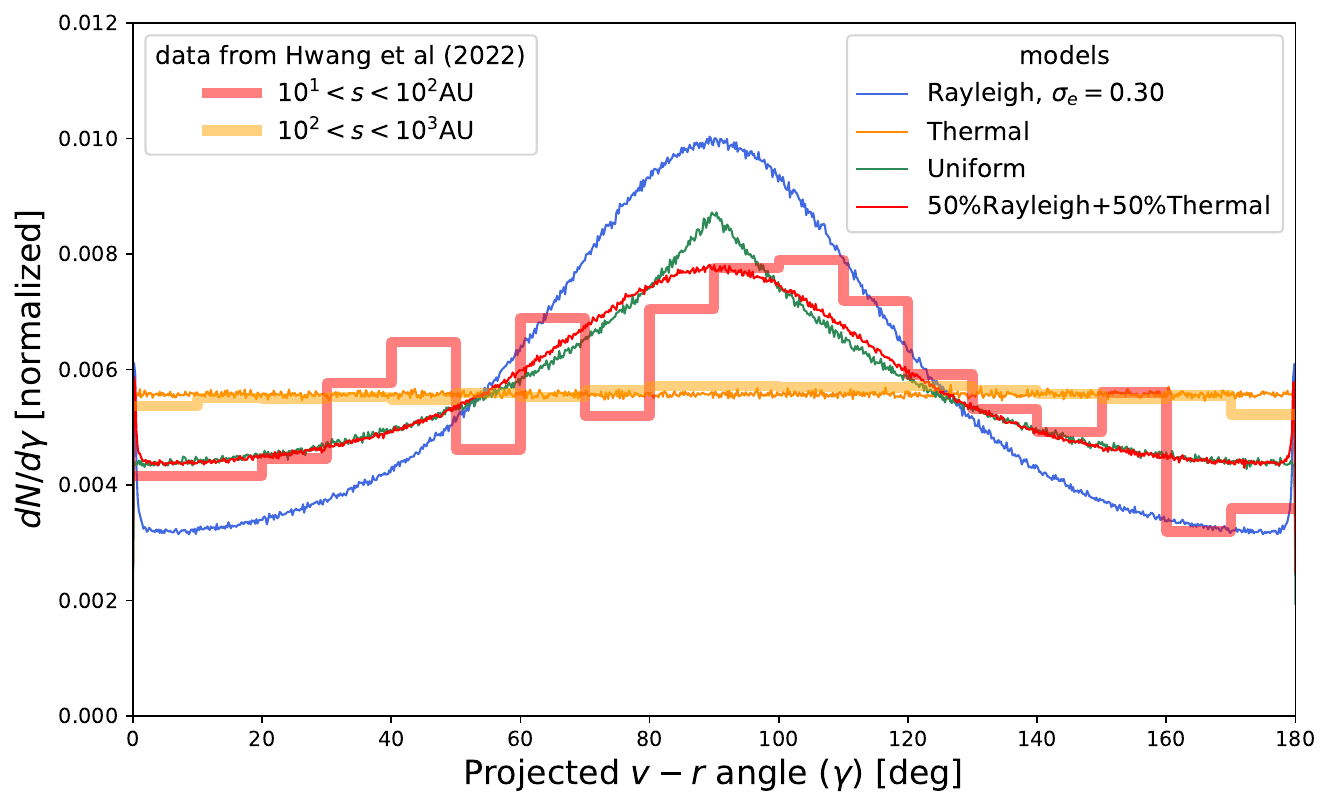}
\caption{Distribution in the  sky-projected displacement-velocity angle $\gamma$ for resolved binaries. The observed values (thick histograms for different values of projected separation $s$)  are adopted from Fig. 6 of \citet{Hwang2022}, and different models are in thin lines, with our best-fit mixture model (\S \ref{sec:model-compare}) in red.
Wide binaries ($10^2 < s < 10^3$au) are best described by a thermal distribution, but close binaries ($10^1 < s < 10^2$au)
can be described equally well by a uniform distribution ($dN/de = {\rm const}$) or by our mixture model. 
}
\label{fig:mixed}
\end{figure}

\section{The unimportance of stellar encounters}
\label{sec:unimportance}

Could the AU-scale binaries have obtained their eccentricities by scattering passing stars, especially while they are still within their birth clusters?

We hold that this is  unlikely. We present multiple arguments. 

For our AU-binaries, we are in the `hard' binary case \citep{Heggie1975}, where the binary orbital velocity ($V_{\rm orb}$) exceeds the mean dispersion velocity of the cluster ($ V_\infty$, e.g, $V_\infty \sim 12$\,km/s in the dense core of 47 Tuc).
If we consider impacts with the closet approach distance $R_{\rm min} \geq a$ (binary separation), the encounter is in the adiabatic limit
where the binary have time to revolve multiple times during one close-approach of the third star. Moreover, the third star orbit is close to being parabolic. An initially circular binary will receive a kick in eccentricity that is of order \citep{Heggie1996,Spurzem2009},
\begin{eqnarray}
\delta e & \approx &
3\sqrt{2\pi} \frac{M_3 M_{12}^{1/4}}{M_{123}^{5/4}} \left(\frac{2 R_{\rm min}}{a}\right)^{3/4}\,\nonumber \\
& & \times \exp\left[- {2\over 3} \left(\frac{2 M_{12}}{M_{123}}\right)^{1/2}
\left(\frac{R_{\rm min}}{a}\right)^{3/2}\right]\, ,
\label{eq:deltae}
\end{eqnarray}
where the total binary mass $M_{12}=M_1+M_2$, and $M_{123}=M_1+M_2+M_3$ with $M_3$ being the perturber mass. For $V_{\rm min} = \sqrt{GM_{12}/R_{\rm min}} \gg V_\infty$, the impact parameter is of order $b \approx R_{\rm min} V_{\rm min}/V_\infty$ (or else $b \sim R_{\rm min}$).   We adopt the set of masses, $M_i = [1.0,0.7,0.3]$ and find $R_{\rm min} \sim 2 a$, if one requires a kick magnitude $\delta e = 0.3$. The adiabatic limit brings about an exponential suppression of the kick. This means we need only to account for the one encounter that has the closest impact parameter. All other encounters do not add substantially. 

Consider a cluster with total mass $M_{\rm c}$ and size $R_{\rm c}$, $V_\infty \sim \sqrt{G M_c/R_c}$, number density of stars $N_c = M_{\rm c}/M_*/\pi R_c^3 $. The mean-free-time to have an impact such that $R_{\rm min} \sim 2a$ is
\begin{eqnarray}
    t & \sim &  \frac{1}{n (\pi b^2) V_\infty} 
    \sim 
    4\times 10^8 {\rm yrs}  
      \left(\frac{N_c}{10^5/{\rm pc}^3}\right)^{-1}
      \left(\frac{V_\infty}{10\km/\s}\right)\,\nonumber \\
      & & \times 
      \left(\frac{M_{12}}{1.7M_\odot}\right)^{-1}
      \left(\frac{a}{1{\rm AU}}\right)^{-1} \, .
 \label{eq:meanfreetime} 
\end{eqnarray}
where we have scaled the cluster density and velocity by values appropriate for a very dense region, the core of 47 Tuc 
\citep{McLaughlin2006}. Most stars are formed in much less dense clusters that dissolves in tens to hundreds of million years. Outside these birth clusters, the density is much lower and the impact is even smaller. 
So overall, AU-scale binaries are unlikely to have been affected by passing-by stars \citep{Parker2009}. 

Even if the above estimate is wrong, we argue that an origin in stellar scattering can be excluded. First, scattering tends to affect the wider binaries more, while the observed Rayleigh mode is largely invariant with the orbital period. Second,  if stellar scattering is so strong as to produce $e\sim 0.3$ for AU-scale binaries, it would also have dissolved all binaries that are a few times wider \citep{Heggie1975,Spurzem2009}. 


\section{Numerical Simulations of Brown Dwarf Ejections}\label{sec:scatterings}

We simulate the gravitational interactions between stellar binary and multiple brown dwarfs, with a focus on the effects on binary eccentricity. 

The binary is composed of a $0.7 M_\odot$ primary and a $0.3 M_\odot$ secondary. They are initialized with a circular orbit with $a=a_{\rm bin,init} = 1$AU. A crowd ($N_{\rm bd}$) of brown dwarfs each with mass $M_{\rm bd}$ is uniformly sprinkled  in logarithmic distance (from $0.03 a_{\rm bin,init}$ to $0.9 a_{\rm bin,init}$) between the binary stars. All initial orbital angles are drawn randomly. The initial eccentricities are zero, and the mutual inclinations are drawn from a Rayleigh distribution with $\sigma_i = 10^{-3}$rad. {\w For our choice of parameters, ejections typically occur before physical collisions do. So} we ignore collisions and set all physical sizes of the particles to zero. 
We integrate the system using the IAS15 integrator in REBOUND \citep{Rein2012,Rein2015}. 

We carry out 100 simulations for each of the following two sets of parameters: $N_{\rm bd}=6$, $M_{\rm bd}=0.05 M_\odot$; $N_{\rm bd}=30$, $M_{\rm bd}=0.01 M_\odot$. The dynamics are similar. {\w The integrations are carried out for tens of thousands of orbits.} Within a short  time (typically tens of orbits), most of the brown dwarfs have been ejected. The remaining few may become bound to one or the other stars. During these ejections, the initial conditions are quickly erased, and we observe that the binaries are hardened and become eccentric. The binary e-distribution is Rayleigh in form, with best-fit modes  $\sigma_e = 0.19$ and $0.35$, for the two cases respectively. {\w We find that the binary plane remains largely unchanged, because the initial orbits are largely aligned and most scatterings occur far from the stars and are 2-D in nature.}

The Rayleigh mode is  sensitive only to the individual brown dwarf mass, as indicated by eq.\ (\ref{eq:equipartition}). The number of brown dwarfs is not relevant. In fact, even scattering as few as one brown dwarf may be sufficient to achieve the same e-distribution, as is observed in the 2-planet scattering experiments of \citet{Ford2008}.



\section{Enough free-floating Brown Dwarfs?}\label{sec:enoughbd}

If close binaries acquire their eccentricities by ejecting brown dwarfs, one expects to see those ejected bodies. 

We crudely estimate their contributions to the Galactic mass function, for two limiting cases. In the first case (`excitation') we allow one brown dwarf per system, enough to excite the observed eccentricity.  In the second case  (`hardening') we assume that all close binaries  were originally wide, but scatter enough brown dwarfs to wind up at their current separations.

Solar-type binaries in the field follow a log-normal  distribution in orbital periods \citep{DM91,Raghavan2010,Moe2019}.
\begin{equation}
{{df_{\rm binary}}\over{d\log P}}\approx {{f_{\rm binary}}\over{\sqrt{2\pi} \sigma_{\log P}}}\, 
\exp\left[ -{1\over 2} \left(\frac{\log P - \log P_0}{\sigma_{\log P}}\right)^2\right]
    \label{eq:Pdist}
\end{equation}
with the total binary fraction $f_{\rm binary} \sim 0.50$, $\log P_0\sim 5$ (or $\sim 40$au),  and  $\sigma_{\log P} \sim 2.3$. Here, all logarithms are 10-based, and all  periods are in unit of days.
{\w Counting binaries inwards of $50$au, we find a total number fraction of $f \approx 0.25$. So if one ejected brown dwarf is responsible for exciting the eccentricity in one binary, we expect a brown dwarf to star ratio of $\sim 1/4$.}



To harden binaries from an initial separation $a_0$ to a new separation $a \ll a_0$, the amount of mass ejected (with parabolic orbits) is $M_{\rm ejected} \sim \ln (a_0/a)\,  M_2  \sim 2/3 \ln (P_0/P) \, M_2$.  Taking $M_2/M_{\rm bd} \sim 10$, the number ratio of brown dwarfs to stars is
\begin{eqnarray}
\frac{N_{\rm bd}}{N_*} 
& \approx &  {2\over 3} \times 10 \times\int^{P_0}_{P=10d} {{df_{\rm binary}}\over{d\log P}} \times (\ln P_0 - \ln P)\, d\log P \nonumber \\ & \approx & 5.5\, .
\label{eq:hardening}
\end{eqnarray}

We compile observational constraints using two different types of studies: those that count objects  in young stellar clusters, and those that employ  microlensing. Using data from {\w Pleiades} (120 Myrs old), \citet{Moraux2003} showed that the mass function at the low end can be well fit by a log-normal distribution (their eq. 3). In even younger clusters ($\sigma$ Ori at 3 Myrs, Upper Sco at 5-10 Myrs), there appears to be more free-floating objects than described by this log-normal form: for masses below $\sim 0.05 M_\odot$ \citep{Lodieu2013,Bejar2011} found a mass-function that roughly goes as $dN/d\log M \propto M^0$. So we append a flat tail to the above log-normal distribution. For the microlensing data, we adopt solution CR1 in \citet{Sumi2023}, inferred using lensing events from the Galactic bulge. These two mass functions differ in form (left panel of Fig. \ref{fig:nomass}), but they give roughly the same {\w results:
the number of brown dwarfs to stars is $\sim 1/3$. This is similar to that inferred by \citet{Kirkpatrick2024} from a population census in the local 20\,pc volume, $1/4$, possibly rising to $1/3$. This means there are enough brown dwarfs to excite the eccentricities, but likely not enough to harden all close-binaries.}



This comparison also suggests that, if the ejection scenario is correct, almost all brown dwarfs should be formed in massive disks, as companions to stellar binaries. 

An interesting prediction for such a scenario is that the ejection event may leave evidence in the kinematics of these bodies. They should be leaving their birth systems with a velocity dispersion that is of order the orbital escape velocity. {\w 
For a 50\,au binary (peak of the binary) with $M_1 = 0.3M_\odot$ (peak of the initial mass function), this is $\sim 2$ km/ s, comparable to the typical dispersion in many young clusters (e.g., Orion). More studies are warranted.}

\begin{figure*}
\centering
\includegraphics[width=0.95\textwidth]{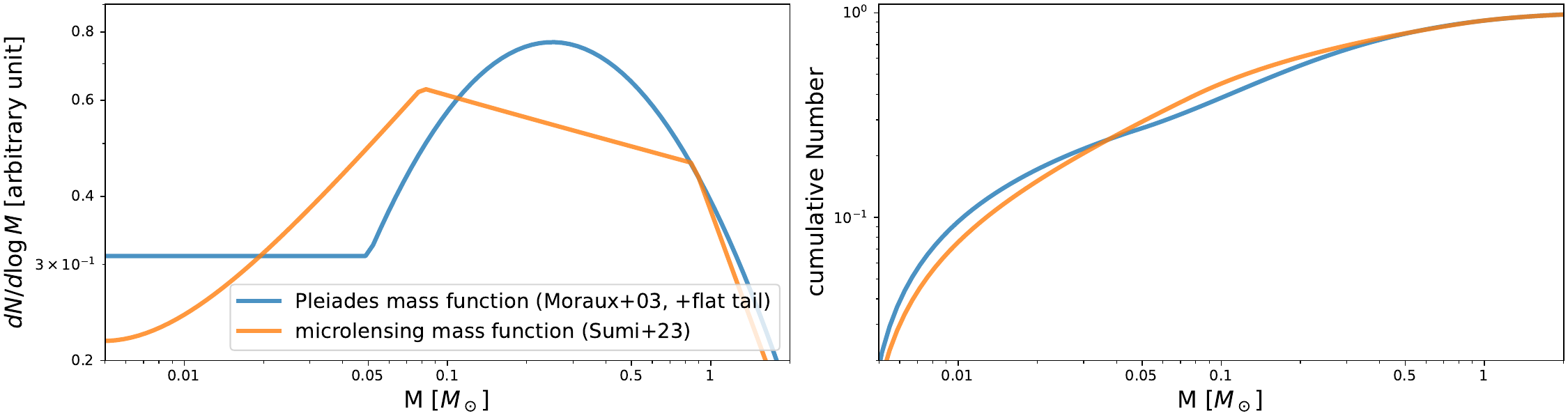}
\caption{Mass functions in the low-mass end, in differential (left panel) and cumulative forms (right panel{\w , cumulative number}). The two determinations are from young clusters and from microlensing of bulge stars. For both cases, 
{\w the number ratio between brown dwarfs (taken to be below $0.05 M_\odot$) and stars is $\sim 1/4$,}
enough to excite Solar-type close binaries 
but not enough to harden them 
}
\label{fig:nomass}
\end{figure*}

\section{Is there only a single Rayleigh?}\label{sec:model-compare}

We have obtained a best-fit Rayleigh of $\sigma_e = 0.30$ for our `gold' sample. But given the large sample size ($N\sim 3000$), it is possible to extract more information. We query the data to determine if it prefers other solutions, in particular,  two separate Rayleigh distributions that may arise when there are  two different physical processes at play. 

In the following, we employ a Bayesian framework to answer this question. 
The marginal likelihood, or evidence, that data $D$ are generated by a model $\mathcal{M}$, is given by the integral
\begin{equation}
    \mathcal{E}(D|\mathcal{M})
    =\int \mathrm{Pr}(\pmb{\theta}|\mathcal{M})\mathrm{Pr}(D|\pmb{\theta},\mathcal{M}) d\pmb{\theta}\, ,
\end{equation}
{\w where $\pmb{\theta}$ is the random variable to be marginalized over.}
The ratio between marginal likelihoods for two different models then gives a Bayes factor that provides one metric for how much more (or less) strongly a given model is supported by the data. In our case the data $D$ are binary eccentricities in the `gold' sample. For a given data set $D=\{e_1,...,e_i,...,e_N\}$ we incorporate observational uncertainty by writing $\mathrm{Pr}(D|\pmb{\theta},\mathcal{M})
=\prod_{i=1}^N\int \mathrm{Pr}(e_i|e_\textrm{true})
\mathrm{Pr}(e_\textrm{true}|\pmb{\theta},\mathcal{M})
\text{d}e_\textrm{true},$ where we adopt $\mathrm{Pr}(e_i|e_\textrm{true})$ to be  a truncated Gaussian  distribution with a standard deviation of $0.025$. 

\subsection{Single Rayleigh}
Model $\mathcal{M}_1$ assumes that the eccentricities of the entire population are distributed according to a single Rayleigh distribution, 
\begin{eqnarray}
    \mathrm{Pr}(D|\pmb{\theta}_e,\mathcal{M}_1) &  = & \prod_{i=1}^N \frac{e_i}{\sigma_e^2}\exp\left(-\frac{e_i^2}{2\sigma_e^2}\right) 
    \nonumber \\
   &  = &\sigma_e^{-2N}\mathcal{P}\exp\left(-\frac{S}{2\sigma_e^2}\right)\, ,
\end{eqnarray}
where $\mathcal{P}=\prod_i^Ne_i,$ and $S=\sum_i^Ne_i^2.$ For a uniform probability distribution $\mathrm{Pr}(\sigma_e|\mathcal{M}_1)=(\Delta\sigma)^{-1}$, where $\Delta\sigma=\sigma_\textrm{max}-\sigma_\textrm{min}$
for $\sigma_\mathrm{min} \leq \sigma_e \leq \sigma_\mathrm{max}$,
the evidence is then given (in the absence of observational uncertainty) as
\begin{align}
   & & \mathcal{E}(D|\mathcal{M}_1)
    =(\Delta\sigma)^{-1}
\int_{\sigma_\mathrm{min}}^{\sigma_\mathrm{max}}
    \mathcal{P}
    \sigma_e^{-2N}\exp\left(
        -\frac{S}{2\sigma_e^2}
    \right)d\sigma_e
\\\notag
    & &=(\Delta\sigma)^{-1}2^{N-3/2} S^{1/2-N}\mathcal{P}
    \Gamma \left(N-1/2,\frac{S}{2 \sigma_e^2}\right)
    \Bigg|^{\sigma_\mathrm{max}}_{\sigma_\mathrm{min}}\, ,
\end{align}
where $\Gamma$ is the incomplete Gamma function. We find that, for our `gold' sample, the likelihood $\textrm{Pr}(D_\textrm{Gaia}|\sigma_e,\mathcal{M})$
approaches a singularly peaked function at $\sigma_e = 0.30$. So $\mathcal{E}(D_\textrm{Gaia}|\mathcal{M}_1)\gg1$ for any choices of $\sigma_\textrm{max}$ and $\sigma_\textrm{min}$, regardless of the observational uncertainties. 

\subsection{Double Rayleigh?}

Can the Gaia sample be instead drawn from two separate Rayleigh distributions? 
We consider a mixture model $\mathcal{M}_2$ with
\begin{eqnarray}
   \mathrm{Pr}(D|\sigma_1,\sigma_2,f,\mathcal{M}_2) 
    & = & \prod_{i=1}^N 
    \left[
        f\dfrac{e_i}{\sigma_1^2}\exp\left(-\dfrac{e_i^2}{2\sigma_1^2}\right)
     \right.\nonumber \\
     & & \left. +(1-f)\dfrac{e_i}{\sigma_2^2}\exp\left(-\dfrac{e_i^2}{2\sigma_2^2}\right)
    \right],
\end{eqnarray}
where $0\leq f\leq1.$ Assuming 
$\textrm{Pr}(\sigma_1,\sigma_2,f|\mathcal{M}_2)
=1/(\Delta \sigma)^2$, we compute the evidence $\mathcal{E}(D_\textrm{Gaia}|\mathcal{M}_2)$ numerically (using Clenshaw-Curtis quadratures). For like values of $\sigma_\textrm{max}$ and $\sigma_\textrm{min},$ we find $\mathcal{E}(D_\textrm{Gaia}|\mathcal{M}_1)/\mathcal{E}(D_\textrm{Gaia}|\mathcal{M}_2)\sim2-5$ when $\sigma_\textrm{min}\gtrsim0.1$ and $\sigma_\textrm{max}>0.303.$ This does not provide particularly strong support for a single Rayleigh distribution over a double one. However, for any intermediate mixing fraction $0<f<1$, we find that the likelihood $\mathcal{L}=\textrm{Pr}(D_\textrm{Gaia}|\sigma_1,\sigma_2,f,\mathcal{M}_2)$ peaks at $\sigma_1\approx\sigma_2\approx0.303$, which simply corresponds to a single Rayleigh distribution with the same mode; the flexibility of the mixture model provides no added utility. 

When $\sigma_\textrm{min}\lesssim0.05$, the Gaia data provide a stronger evidence for a double Rayleigh model with $\sigma_1=0.303,$ $\sigma_2\ll1$, and $1-f\ll1$ (i.e., mostly the primary Rayleigh mode $\sigma_1=0.303$). However, we discount this model for two reasons. First, it amounts to an over-fit of the small eccentricity data. Second, there is a strange (but small) excess of $e=0$ points in the data (as seen in the left panel of Fig. \ref{fig:x_extrapolate}) that likely reflects issues in the Gaia pipeline.

}

\end{appendix}

\end{document}